\documentstyle[12pt]{article}

\begin{document}

\author{G.V.~Vlasov \\
Landau Institute for Theoretical Physics and \\
Moscow Aviation Institute, \\
2 Kosygin Street 117334, Moscow, Russia\thanks{
E-mail: vs@itp.ac.ru}}
\title{Generalized fluid dynamics and the boundary condition}
\maketitle

\begin{abstract}
The purpose of the present paper is to work out the general formulation of
the relativistic fluid dynamics with vorticity (including relativistic
superfluid) on the total manifold with boundary. Making use of the Hodge
decomposition, we emphasize that the generalized equations of motion include
a term due to non-trivial topology, while the pure vortex term introduced by
Carter and Langlois \cite{CL} will be presented merely by a co-exact form.
We also consider a vortex sheet combined of individual vortices and discuss
the boundary problem.
\end{abstract}

\sloppy

{\bf 1.} The variational procedure yielding the relativistic fluid equations
of motion can be discussed in diverse ways \cite{Israel,CK,EKHMR}. We begin
with a formulation which may be related both to the fluids and fields.

Consider $n$-dimensional manifold $X$ without boundary. For an exterior
$p$-form $\alpha _{\nu _1...\nu _p}$ we define the exterior derivative $D$
according to the formula~\cite{CDWD,Schutz}

\begin{equation}
D\,\alpha \,=(\,p\,+1)\,\nabla _{[\,\varrho \,}\alpha _{\nu _1...\nu _p]}
\label{extder}
\end{equation}
If we deal also with vectors, we can determine for an arbitrary $k$-vector $%
\vec a\equiv a^{\nu _1...\nu _\kappa }$ its divergence as~\cite{Schutz}
\begin{equation}
\delta \,\vec a=(-1)^{n(k-1)}\,\,^{*}(D\/\/\/_{*}\vec a)=\nabla _{\nu
_{k\,\,}}a^{\nu _1...\nu _k}  \label{diver}
\end{equation}
where the star operator $^{*}$ transforms the $p$-forms into the dual $(n-p)$%
-vectors, while $_{*}$ makes the inverse mapping, namely

\begin{equation}
^{\ast }\alpha =\frac 1{p!}\,\epsilon ^{\mu _1...\mu _n}\alpha _{\mu
_1...\mu _p}=\vec \epsilon \cdot \alpha \qquad \qquad _{*}a=\frac
1{q!}\,\epsilon _{\mu _1...\mu _n}a^{\mu _1...\mu _p}=\epsilon \cdot \vec a
\label{dual}
\end{equation}
Here a dot denotes the inner product, $\epsilon ^{\/\mu _1...\mu _n}$ is a
fully antisymmetric tensor. Note that $\vec \epsilon \cdot {\bf \epsilon }=n$
and~\cite{Schutz}~$\quad ^{*\,}{}_{*}\,\vec a=\left( -1\right) ^{q(n-q)}$%
\thinspace $\vec a$ . The divergence of an exterior $p$-form is determined~as
\cite{CDWD}~$\quad \delta \,{\bf \alpha }=\left( -1\right) ^{np+p+1}*\left(
D\,*\!\alpha \right) $ , where the Hodge operator $*$ maps the $p$-forms
onto the dual $(n-p)$-forms~$\quad *{\bf \alpha }=\frac 1{p!}\,\epsilon
_{\nu _{(n-p+1)}...\nu _n}^{\,\,\nu _1\,...\,\nu _p}\alpha _{\nu _1...\nu _p}
$ , with the indices risen by the metric tensor $g^{\mu \nu }$.

The Lie derivative $\pounds _\xi $ of an exterior form $\alpha $ along the
vector field $\vec \xi $ obeys the Cartan formula~\cite{CDWD,Schutz}
\begin{equation}
\pounds _\xi \,\alpha =D\/\left( \vec \xi \cdot \alpha \right) +\vec \xi
\cdot D\,\alpha  \label{Lie}
\end{equation}

{\bf 2.} First of all, let us obtain equations of the covariant fluid
mechanics on a $n$-dimensional space-time manifold $X$ without boundary, $%
\partial X=0$. Following Carter and Langlois \cite{Carter,CL}, consider a
fibre bundle $(X,Y,g)$ with projection $g:X\rightarrow Y$ which maps the
total space-time with the local coordinates $x^\mu $ onto $(n-1)$%
-dimensional base manifold $Y$ with local coordinates $y^a$ where $%
a=1,...,(n-1)$. A $1$-dimensional fiber $F_y$ under the point $y\in Y$ can
be interpreted as a streamline passing at this point. If we introduce a $%
(n-1)$-form $N_{a_1...a_{(n-1)}}$ closed on the manifold $Y$, the
contra-variant functor $g^{\star }:\,\Omega \left( Y\right) \rightarrow
\Omega \left( X\right) $ determines a closed $(n-1)$-form $N\equiv N_{\nu
_1...\nu _{(n-1)}}=g^{\star }\left( N_{a_1...a_{(n-1)}}\right) $ on $X$ ~%
\cite{CDWD}. The particle number current $\vec n$ is introduced, in light of
(\ref{dual}), as a dual to $N$. The closure condition $DN=0$, according to (%
\ref{extder}), (\ref{diver}) leads to the conservation law $\nabla _\varrho
\,n^\varrho =0$.

The dynamical equations are derived from the variational formula for the
Lagrangian
\begin{equation}
d\,L=\frac 1{\left( n\,-1\right) !}\,H^{\mu \nu ...\varrho }\,dN_{\mu \nu
...\varrho }\,+\frac 12\lambda ^{\mu \nu }\,\,d\,w_{\mu \nu }
\label{lagrange}
\end{equation}
where the variable $\vec H=\,^{*}\mu $ conjugated to $N$ is regarded as the
momentum vector. (We omit the terms due to the variations of metric since
they lead to the Einstein equations in matter not discussed here). As soon
as quantum vortices exist in superfluid, the irrotationality condition $%
D\,\mu =0$ breaks down, and the Lagrangian depends on the closed vorticity $%
2 $-form $w=D\,\mu $ .

Consider a fibre bundle $\left( X,Z,\,p\right) $ with the projection $%
h:X\rightarrow Z$ of the total manifold onto $2$-dimensional base $Z$ with
local coordinates $z^A\quad $($A=1,2$). For the two-form $w_{AB}$ closed on $%
Z$ the contravariant functor $h^{\star }:$ $\Omega ^2\left( Z\right)
\rightarrow \Omega ^2\left( X\right) $ determines the vorticity $2$-form $%
w\equiv w_{\varrho \nu }$ closed on $X$. The two-dimensional fibres are
interpreted as world sheets of the quantum vortices.

Carter and Langlois~\cite{CL} present the particle number form through a
Kalb-Ramond field $N=D\,B$, that, in view of~(\ref{diver}),~(\ref{dual}),
is equivalent to $n^\varrho =\nabla _\nu \,b^{\varrho \,\nu }$, where the
bi-vector $\vec b$ is the dual to $B$. However such a representation is not
global for a manifold $X$ with the Betti number {\bf {b}$^{n-1\,}\left(
X\right) \neq 0$ }that, indeed, takes place when quantum vortices appear:
the closed form $N$, according to the Poincar\'e lemma~\cite{CDWD}, is exact
only on a manifold homeomorphic to a sphere. In the general case one should
write $N=F+D\,B$, or $\vec n=\vec f+\delta \vec b$, where $(n-1)$-form $F$
belongs to the cohomology group $H^{\left( n-1\right) }\left( X,\,\Re
\right)$.

Variations of dynamical variables in Eq.~(\ref{lagrange}) are not arbitrary,
they obey the convective variational principle. For an infinitesimal
displacement of the coordinates $\delta \,y^a=-\,\xi ^\varrho \,\left(
\partial y\/^a/\partial \,x^\varrho \right) $ and $\delta \/z^A=-\,\xi
^\varrho \,\left( \partial z^A/\partial \,x^\varrho \right) $ along an
arbitrary vector field $\vec \xi $ possible variations of the closed
forms
$N$ and $w$ are given through the corresponding Lie derivatives~(\ref{Lie})
as
\begin{equation}
dN=\pounds _\xi \,N=D\left( \vec \xi \cdot N\right)  \label{nlie}
\end{equation}
\begin{equation}
dw=\pounds _\xi \,w=D\left( \vec \xi \cdot w\right)  \label{wlie}
\end{equation}
Then, substituting eqs.~(\ref{nlie}),~(\ref{wlie}) in~(\ref{lagrange}) and
integrating over $X$, we obtain the first variation of the action integral
\begin{equation}
\begin{array}{c}
d\,S=\int\limits_XdL=\langle D\left( \vec \xi \cdot N\right) \mid \vec
H\rangle +\langle D\left( \vec \xi \cdot w\right) \mid \vec \lambda \rangle =
\\
\left( -1\right) ^{n-1}\,\langle \vec \xi \cdot F\mid \delta \vec H\rangle
+\left( -1\right) ^{n-1}\,\langle \vec \xi \cdot DB\mid \delta \vec H\rangle
+\left( -1\right) ^{3n-1}\,\langle \vec \xi \cdot w\mid \delta \vec \lambda
\rangle
\end{array}
\label{action}
\end{equation}
Let us introduce a $(n-2)$-form $L=\,_{*}\lambda $,and present it as an
orthogonal sum of a harmonic $G$, exact $E$ and co-exact form $C=\delta Q$
by the Hodge decomposition~\cite{Schwarz}:
\begin{equation}
L=G+E+C  \label{decompose}
\end{equation}
Since vector $\vec \xi $ is arbitrary, the eq.~(\ref{action}), in view of~(%
\ref{nlie}),~(\ref{diver}) and~(\ref{dual}) determines the equation of motion

\[
\left\{ \,\vec f+\delta \vec b+\left( -1\right) ^{n+1}\,\delta \vec
c\right\} \cdot w=0\qquad \qquad \vec c=\,^{*}C
\]
or
\begin{equation}  \label{motion}
\left\{ \,f_\nu +\nabla _\sigma b_\nu ^\sigma +\left( -1\right)
^{n+1}\,\nabla ^{\sigma \,}\nabla _{[\sigma }\,\psi _{\nu ]}\right\}
\,w_\varrho ^\nu =0\qquad \qquad \psi =*Q
\end{equation}
In accordance to the isomorphism between the harmonic forms and the
cohomology groups~\cite{CDWD}, we may seek, without loss of generality, the
solution as a harmonic form $\Delta F=\left( D\,\delta +\delta D\right) F=0$
. Since $f_\nu =\left( *F\right) _\nu $ and~\cite{CDWD} $*\Delta F=\Delta *F$%
, therefore
\begin{equation}  \label{garmon}
\nabla _{\varrho \,}\nabla ^\varrho \,f_\nu \,+R_\nu ^\varrho \,f_\varrho =0
\end{equation}
where $R_\nu ^\varrho $ is the Ricci tensor of the manifold~$X$.

So, the complete problem includes the solving of the dynamical equation ~(%
\ref{motion}) together with eq.~(\ref{garmon}). Carter and Langlois~\cite{CL}
have obtained an equation analogous to ~(\ref{motion}) on a $4$-dimensional
total manifold with the relevant Betti number being equal to zero. Now we
conceive that the presence of quantum vortices is described by $1$-form $%
\psi $ instead of $2$-vector $\vec \lambda $ --- or,~that~eq.~(\ref{motion})
involves only the co-exact part of $(n-2)$-form~(\ref{decompose})~---~,
while the additional topological term ~$f$ may exist even when the vortices
are absent.

{\bf 3.} The world-sheet of quantum vortices correspond to a $2$-dimensional
support {\sc supp}$\,w$ of the vorticity $2$-form and, hence, to a $2$%
-dimensional base-manifold $Z$. In principle, {\sc supp}$\,w$ can be a $3$%
-dimensional manifold - an analog of the so called vortex sheet (do not mix
this term with a world-sheet!) in superfluid He-3~\cite{PTVet}. For $n$%
-dimensional total manifold one may suggest {\sc supp}$\,w$ to be also a $p$%
-dimensional membrane ($(2<p<n)$); this time the base $Z$ will be a $p$%
-dimensional manifold. Since the vortex sheet is combined of individual
vortices it must be an orientable manifold; the M\"obius band, for instance,
cannot be a vortex sheet. In other words a vortex membrane must be presented
as a direct product of a $(p-2)$-dimensional manifold and a $2$-dimensional
vortex world sheet. Without the loss of generality we can take the base
manifold in the form $Z=S^{p-2}\times \sum $, where $S^{p-2}$ is a $(p-2)$%
-dimensional sphere. Consider a spherical bundle $\left( Z,\sum ,\,\pi
\right) $ with the canonical projection $\pi :Z\rightarrow \sum $ of the
manifold $Z$ with local coordinates $z^A\quad (A=1,...,p-2)$ onto $2$%
-dimensional base $\Sigma $ with local coordinates $\sigma ^u\quad (u=1,2)$.
For an exterior $2$-form $w_{uv}$ closed on $\sum $ the reciprocal image $%
\Omega ^2\left( \sum \right) \rightarrow \Omega ^2\left( Z\right) $
determines the $2$-form $w_{AB}$ closed on $Z$. On the other hand, we can
consider a fibre bundle $\left( Z,S^{p-2},\,\tilde \pi \right) $ where the
vortex world sheet $\sum $ will play the role of a fibre. Thus the closed $2$%
-form $w_{AB}$ is found; the rest procedure is similar to that discussed
above.

{\bf 4.} Now we consider the role of boundary. 

If the manifold $X$ has the boundary $\partial X$, we introduce the
inclusion $j:\,X\hookrightarrow \partial X$ and a contravariant functor $%
j^{*}:\,\Omega (\partial X)\rightarrow \Omega (X)$ . The tangential
component of form $\alpha $ is determined as $\alpha ^{\parallel
}=j^{*}\alpha $, while $\alpha ^{\perp }=\alpha -\alpha ^{\parallel }\,$ is
regarded as the normal component. Integration over the total manifold with
boundary, satisfies the formula~\cite{Schwarz} 
\begin{equation}
\langle D\alpha ,\beta \rangle _X=\langle \alpha ,\delta \beta \rangle
_X+\langle \,\alpha ^{\parallel },\beta ^{\perp }\rangle _{\partial X}
\label{intgr}
\end{equation}
In view of formula~(\ref{intgr}) the equation~(\ref{action}) will contain
additional boundary terms due to the last integral in~(\ref{intgr}). If we
request the equations of motion to be the same as on a boundary-less
manifold, these additional terms must vanish under the appropriate boundary
conditions. Let the Betti number be {\bf b}$^{n-1\,}\left( X\right) =0$ ,
that is $N=DB$. Since the exact and co-exact forms belong to the Dirichlet
and Neumann field respectively~\cite{Schwarz}, we immediately write
\begin{equation}
(DB{}{}\,)^{\parallel }=0  \label{parll}
\end{equation}
\begin{equation}
w^{\parallel }=0\qquad \qquad \left\{ \nabla _{[\nu }\,\psi _{\varrho
]}\right\} ^{\perp }=0  \label{wpsiparll}
\end{equation}
that in view of~(\ref{intgr}) implies no additional terms to appear in Eq.~(%
\ref{action}), and, hence, the equation of motion~(\ref{motion}) remains
valid. When {\bf b}$^{n-1\,}\left( X\right) \neq 0$ , the particle number
form $N$ is not exact, and the extraordinary boundary term 
\[
\langle \,N^{\parallel },(*\mu )^{\perp }\rangle _{\partial X}=\langle
\,*n^{\perp },(*\mu )^{\perp }\rangle _{\partial X}=0 
\]
under the boundary condition 
\begin{equation}
N^{\parallel }=n^{\perp }=0  \label{bound}
\end{equation}
similar to~(\ref{parll}), makes no changes in eq.~(\ref{motion}).

In the non-relativistic limit we substitute $n_\nu \rightarrow n\left( 1,%
{\bf \vec v}_s\right) $ , $w_{0i}=0$ , $w_{ij}\rightarrow {}${\sc rot}$\,%
{\bf \vec v}_s$ , where ${\bf \vec v}_s$ is the superfluid velocity. Then,
the conditions~(\ref{bound}) and~(\ref{wpsiparll}) reduce respectively
to $\vec v_s^{\,\perp }\,=0\,$
and $\left( {\sc rot}\,\vec v_s\right)^{\parallel }=0$, implying
the absence of flow into the walls of vessel
containing superfluid and the continuity of vortex lines (i.e. that they are
pinned to the walls or form closed rings but not broken up inside the fluid).

\end{document}